\documentclass{amsart}

\usepackage{amsmath,amsfonts,amssymb,amsthm,dutchcal}
\usepackage{graphicx,xcolor}

\newcommand{\dd}{\mathcal{d}}

\begin{document}

\title{Aspects of Defect Topology in Smectic Liquid Crystals}
\author{Thomas Machon}
\address[TM]{H. H. Wills Physics Laboratory, University of Bristol, Bristol BS8 1TL, UK}
\email{t.machon@bristol.ac.uk}
\author{Hillel Aharoni}
\author{Yichen Hu}
\author{Randall D. Kamien}
\address[HA, YH and RDK]{Department of Physics and Astronomy, University of Pennsylvania, 209 South 33rd Street, Philadelphia, Pennsylvania 19104, USA}

\begin{abstract}
We study the topology of smectic defects in two and three dimensions. We give a topological classification of smectic point defects and disclination lines in three dimensions. In addition we describe the combination rules for smectic point defects in two and three dimensions, showing how the broken translational symmetry of the smectic confers a path dependence on the result of defect addition. \hfill\break\break

\centerline{\sl Dedicated to Maurice Kl\'eman on the occasion of his 84$^{\rm th}$ birthday}
\end{abstract}

\maketitle

\section{Introduction}

Smectic-A liquid crystals (smectics) are materials with one-dimensional positional order (layers) as well as apolar orientational order aligned along the layer normals~\cite{deGennesLC}. Following de Gennes~\cite{deGennes72}, they are typically described in terms of a complex order-parameter, $\psi$, the magnitude of $\psi$ describes the degree of smectic order, and the level sets of phase $\phi=\textrm{Arg}\;\psi$ determine the smectic layers. The additional symmetries of the smectic mean the layers are unoriented, a smectic configuration described by $\psi$ is locally equivalent to one described by its conjugate $\overline{\psi}$ so in general the smectic may be modelled by an unordered pair $\Psi =\{ \psi, \overline{\psi} \}$ of complex-valued scalar fields~\cite{chen09} in a domain $\Omega$ which we will take to be Euclidean $\mathbb{R}^2$ or $\mathbb{R}^3$. $\Psi$ may be thought of as a complex scalar field with a $\mathbb{Z}_2$ gauge symmetry. $\Psi$ determines the local density of the smectic, given as
\begin{equation}
\rho = \rho_0 + \delta \rho \cos (\phi) = \rho_0+\textrm{Re}[\psi],
\end{equation}
with the smectic layers located at maxima of density. The free energy of the smectic favours a fixed value of $|\psi |$ as well as a fixed wavelength for the density oscillations, so that in Euclidean space the smectic groundstates may be written as double valued plane waves
\begin{equation}\Psi_{gs} = |\psi_0| \left \{  e^{i(\bf q \cdot \bf r + \phi_0)}, e^{-i(\bf {q} \cdot \bf {r} + \phi_0)} \right \},
\label{eq:groundstates}
\end{equation} 
where the layers (regions of maximum density where  $\cos ( {\bf q} \cdot {\bf r} + \phi_0) =1$) are equidistant codimension-one hyperplanes. The values of $|\psi_0|$ and $| {\bf q}|$ are fixed by material parameters, so the space of groundstates is determined by $\phi_0$ and ${\bf n}={\bf q}/|{\bf q}|$, which determine the density at the origin and layer orientation respectively.

In physical systems there are regions where the smectic order breaks down, forming singularities. According to the standard homotopy theoretic approach~\cite{mermin79}, the possible topological classes of defects are determined by the homotopy groups of the groundstate manifold, $E_d$. Given \eqref{eq:groundstates}, $E_d$ is equivalent to the space of equidistant unoriented codimension-one hyperplanes in Euclidean space and can be described by the pair $({\bf n},\phi_0)$, where $\bf n \in S^{d-1} \subset \mathbb{R}^{d}$ and $e^{i \phi_0} \in S^1 \subset \mathbb{C}$, along with the equivalence relation 
\begin{equation}
({\bf n},\phi_0) \sim (-{\bf n},-\phi_0).
\end{equation}
This relation acts as the antipodal map on $S^{d-1}$ so that ${\bf n}$ lives in real projective space $\mathbb{RP}^{d-1}$. Accounting for the phase variable gives $E_d$ the structure of a twisted circle bundle over $\mathbb{RP}^{d-1}$. For $d=2$ this is a copy of the Klein bottle. For $d=3$ it is a circle bundle over $\mathbb{RP}^2$ such that the union of fibers over each non-contractible great semicircle in $\mathbb{RP}^2$ forms a Klein bottle.

In the classic approach~\cite{mermin79}, codimension-$m$ defects in this system are then classified by homotopy classes of maps $S^{m-1} \to E_d$, $m < d$, and nonsingular textures are described by homotopy classes of maps $S^d \to E_d$. This approach fails as the orientation of the smectic layers is coupled to the gradients of $\phi$, supplying an integrability condition on allowed paths in $E_d$. As a simple example, if $U$ is a simply-connected defect-free region of the smectic, then we obtain the compatibility condition
\begin{equation}
\nabla \phi - |\nabla \phi | \bf n =0.
\label{eq:consist}
\end{equation}
So that arbitrary continuous maps $U \to E_d$ are not permitted. For example, two configurations which are homotopic as maps $U \to E_d$ may fail to be so when demanding \eqref{eq:consist} be satisfied. Perhaps the most notable example of this comes in the theory of two-dimensional crystals: dislocations can be decomposed into disclination dipoles even though the dipole has zero net charge~\cite{kleman_book}.  In smectics, this leads to a subtle distinction between a pincement and a dislocation -- both with disclination dipoles but only the latter has a dislocation.

This failure has been recognised for some time~\cite{mermin79,trebin82}. A conceptually clear but (currently) computationally intractable resolution to this issue is to restrict to maps satisfying the generalisation of \eqref{eq:consist}~\cite{chenthesis}. More concrete progress has been made only in specific cases. In the early 1980s, Po\'{e}naru~\cite{poenaru81} observed that disclinations in smectics can be identified with singularities in measured foliations, and used this to prove that in two dimensional smectics, the winding number, $q$, of any point defect must satisfy the inequality
\begin{equation}
q \leq 1.
\end{equation}
More recently, further work~\cite{chen09} in this area created surface models of two dimensional smectics that clarified the defect topology in two dimensions without the need to add compatibility conditions. In this paper we extend this program, as established by Po\'{e}naru~\cite{poenaru81}. We derive a topological model of smectics, based on the regions of density maxima (`layers') and minima (`half-layers'). We use this to classify topological defects in two and three dimensional smectics. Our work reproduces the known results in two dimensions and gives new results for point and line defects in three dimensions, classifying point defects with graphical trees and line defects with one-loop graphs. Our approach to defect classification is in line with similar previous work~\cite{chen09,trebin82} and can be thought of (at least for point defects) in terms of topological classification of critical points~\cite{king78,king80}. This approach loses the purely algebraic sense of the classical theory and requires a more geometric approach, which appears inherent to the study of systems with broken translational symmetry. 

Furthermore, our approach allows us to construct the `defect algebra', the combination rules that govern the behaviour of smectic defects. In particular, we show that the broken translational symmetry of the smectic leads to an essential path dependence for defect combination rules. This is similar in form to the case of translationally invariant systems whose groundstate manifold possesses a non-Abelian fundamental group~\cite{poenaru77}, but its origins are distinct. The path dependence exists only for `transverse' combinations of defects, where the defects are combined along paths transverse to the smectic layers. The distinction between transverse and the complementary tangential combinations defines a dichotomy in the addition of smectic defects with manifestation of this path dependence being through the creation of additional defects. In particular, we show that point defects in three dimensional smectics are not closed under addition, and necessitate the creation of line defects. We note that in contrast with our defect classification the results on defect addition carry a much more algebraic flavour and have the potential to be systematised in a more general way than presented here.

\section{Smectics and their Symmetries}
\label{sec:2.1}

The smectic energy is minimised by non-zero $\Psi$ with a fixed magnitude and prescribed wavelength. Correspondingly, defects in a smectic may be split into two types: phase singularities (associated to dislocations), where the smectic order melts so that $\Psi =0$ and disclinations, where the layer normals (associated to the gradients of $\Psi$) are singular. The $\mathbb{Z}_2$ symmetry of the smectic ensures that phase singularities are not topologically protected, and may always be broken up into disclinations. This is well-known in the physics literature for the case of edge dislocations, but is true in general, applying also to screw dislocations, for example~\cite{aharoni17,meyer10,achard05, klemenSM, williams75}. The double cover $\Psi$ amounts to describing the smectic as a function $\zeta(x)$ in the upper complex half plane, for all $x \in \Omega$.  Dislocations, phase singularities, only occur at $\zeta^{-1}(0)\subset\Omega$. 
For a sufficiently small neighborhood $U_\dd$ around a defect $\dd$, we define $\chi(x) = \zeta(x) + i \epsilon f(x)$, where $f:\Omega \to [0,1]$ is a smooth function with support in $U_\dd$, and $f(\dd)=1$. Outside $U_\dd$, $\chi=\zeta$  yet, for $\epsilon$ small enough, we have removed the phase singularity without introducing a new one.  It follows that there must be singularities in gradients of $\Phi$: disclinations, singularities in the layer normal.

Because phase singularities can be removed, we will only consider disclinations in the following. Away from such singularities, the normal to the smectic layers is defined by the director field, 
\begin{equation}
N = \{ {\bf n}, - {\bf n} \}.
\label{eq:rel}
\end{equation}
$N$, the layer normal, can be defined for $x\in\Omega$ for which an open neighbourhood $U\subset\Omega$ of $x$ admits a local ordering of $\Phi$.  On $U$ we can choose one of the pair in $\Phi$, $\tilde\phi$ and the unit layer normal is defined as ${\bf n}=\nabla\tilde\phi/\vert\nabla\tilde\phi\vert$ and $N=\{{\bf n}, -{\bf n}\}$ becomes an ordered pair.  The defect set, $\mathcal{D}$, is the union of points and lines in $\Omega$ where $N$ is not defined.  From a topological perspective, this can occur in two ways, either $\Phi$ does not admit a local ordering in a neighbourhood of the defect, or the defect is a simple gradient zero (or singularity). Correspondingly we write $\mathcal{D} = \mathcal{D}_u \cup \mathcal{D}_o$, where $\mathcal{D}_u$ are the non-orientable defects where $\Phi$ does not admit a local ordering ({\it e.g.}, point defects of half-integer winding in two dimensions), and $\mathcal{D}_o$ are the orientable defects.  Note that points in $\dd\in\mathcal{D}_o$ must have zeroes in $\Psi$.  Since we can define an ordered set of phase fields $\Phi$, any non-zero phase winding around $\dd$ implies that $\Psi$ is also defined in an open set around $\dd$ -- it follows that $\Psi$ must have a zero.  Because of this, we note that when we remove phase singularities we must always produce at least some non-orientable disclinations. Again, we note that the standard case of a $+1$ dislocation breaking into two non-orientable disclinations is well-known in the physics literature for the case of edge dislocations, but is true in general.

Given a basis for the tangent space of $\Omega$, $N$ defines a map $\Omega \setminus \mathcal{D} \to \mathbb{RP}^2$, so that smectic defects must obey the standard rules for nematic defects~\cite{mermin79,alexander12}. In other words, one may forget about the layered structure of the smectic and reduce the configuration to that of a nematic. In particular, non-orientable defects must be line-like in three dimensions (in general they may meet at vertices of even degree). 

In a uniform nematic state, the phase field enjoys a global shift invariance, $\Phi\rightarrow \Phi + \{c,-c\}$ for $c\in\mathbb{R}$.  However, the possibility of unorientable defects constrains this symmetry. Because there are no dislocations, $\Phi$ is globally defined and we may evaluate it on the disclinations. If $x \in \mathcal{D}_u$, then the symmetries of the smectic dictate that $\Phi(x)$ must be a fixed point of both the $\mathbb{Z}_2$ symmetry $(\{ 0,0\}$ or $\{\pi, - \pi\})$ and the shift symmetry $\Phi\rightarrow\Phi+\{2\pi,-2\pi\}$. This requires that the defects sit on the level sets $\cos\phi=\pm 1$.

\section{Constructing the Model}

\S\ref{sec:2.1} contained two primary observations. Firstly, phase singularities in smectics are not topologically protected and may always be decomposed into disclinations. Secondly, by symmetry non-orientable defects in smectics must occur where the smectic density is either at a maximum or a minimum. These two facts will inform our topological model of the smectic, which we construct in this section. 

We continue to only consider disclinations. Additionally, we set $ | \Psi | $ to its equilibrium value throughout the system so that the smectic is completely determined by $\Phi$. The smectic density must be at either a minimum or a maximum on non-orientable defects. For our construction we make the assumption that this holds for all defects, including $\mathcal{D}_o$ -- this is consistent with the physical nature of smectics as layered systems, so that defects are only located between layers (density minima) or in layers (density maxima), \textit{i.e.}, that the smectic layers are well defined. While the standard theories of smectics do not distinguish between disclinations at different values of $\phi$~\cite{deGennesLC} (so that there is no Peierls-Nabarro barrier), more detailed theories~\cite{pevnyi14} break this symmetry, giving an energetic preference for defects to lie at specific densities. 

As we described, our construction is to take the sets of points where $\cos \phi=\pm 1$ as a model of the smectic topology. Since all singular points of the smectic are contained on these sets, which we will write as $L_{-1}$ (the `half-layers', regions of density minima) and $L_1$ (the `layers', regions of density maxima). With minor assumptions on the properties of the texture, in particular that the size of regions where $|\nabla \phi| < \epsilon$ are bounded for sufficiently small $\epsilon$, one may adapt standard results~\cite{milnorMorse} to give a decomposition of the gaps between the layers in terms of generalised strips, $\Omega \setminus (L_1 \sqcup L_{-1})=\sqcup S_i$. These strips have the form \begin{equation}S_i \cong M_i\times (0,1),\end{equation} where $M_i$ is an orientable codimension-1 manifold without boundary and $\partial S_i = \mathcal{B}_{i,1} \sqcup \mathcal{B}_{i,-1}$ with each boundary component $\mathcal{B}_{i,j}$ contained in $L_{j}$ (it is possible for two distinct points of $\mathcal{B}_{i,j}$ to correspond to the same point of $L_{j}$, as in the case of the layer structure in the vicinity of a $+1/2$ defect). 
The assumption made here, that the gradient cannot become very small over very large regions, is natural in the context of the smectic free energy which may be written in a defect-free region $U\subseteq \Omega$ as
\begin{equation}
F=\frac{B}{8}\int_U \left [ (|\nabla \phi|^2-q^2)^2 + 16 \lambda^2 H^2 \right]
\label{eq:energy}
\end{equation}
where $H$ is the mean curvature of the layers and $B$ and $\lambda$ are material parameters. The free energy \eqref{eq:energy} penalises deviations from the preferred layer separation, ensuring that $|\nabla \phi|$ is typically $\sim q$. Given this decomposition, one can reconstruct the phase field $\phi$ in each strip, up to deformations that do not affect the topology of the texture. In this way, the sets $(L_{1},L_{-1})$ capture the complete topological information in the texture, in much the same way that the Pontryagin-Thom construction can be used to capture the global topological information in a nematic texture~\cite{chen13}. 

\begin{figure}
\begin{center}
\includegraphics[scale=0.6]{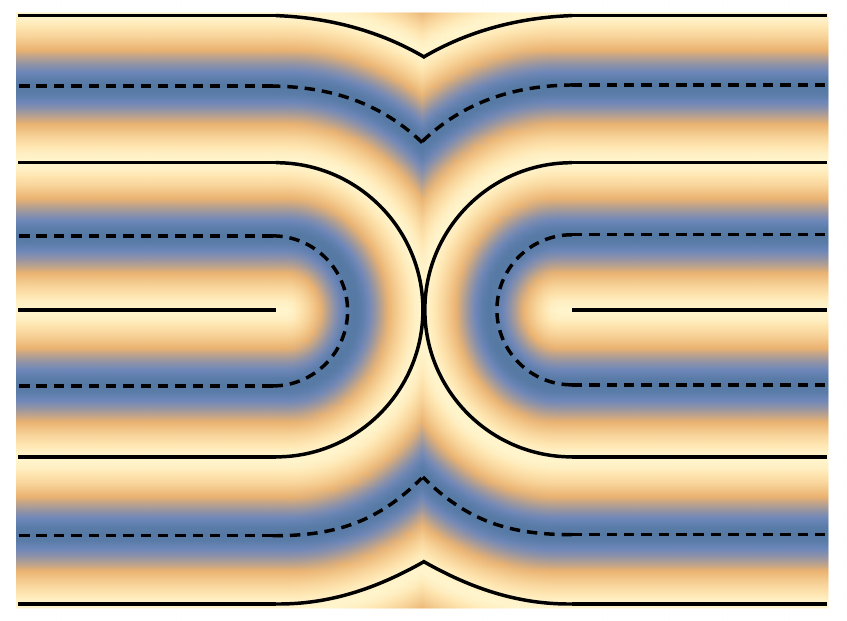}
\end{center}
\caption{Going from a grain boundary smectic configuration to the representation of a smectic textures discussed in the paper. The maxima of the density field $\cos \phi$ are shown by thick black lines and correspond to $L_1$. The minima of density correspond to $L_{-1}$ and are shown by the dashed lines. The configuration contains three disclinations, two $+1/2$ and one $-1$, which correspond to singularities in the layer structure. Note that the gaps between the sets $L_1$ and $L_{-1}$ all have the structure of strips (as discussed in the text), enabling a reconstruction of the texture from the pair $(L_1,L_{-1})$.}
\label{fig:model}
\end{figure}

\section{Structure and Classification of Singularities in Smectics}
\label{sec:classification}

Defects in our system correspond to singularities in the layer structure $(L_{-1}, L_1)$, {\it i.e.}, the points where $L_i$ does not look like a codimension-1 manifold. The topological structure of the defect is therefore determined by the manner in which the smectic layers or half-layers are attached to the defect itself. To probe this, consider a defect $\dd$, defined as a connected component of $\mathcal{D}$, on $L_i$, and let $U_\dd$ be an open neighbourhood of $\dd$ in $\Omega$.  As we shrink the open neighbourhood around $\dd$, the intersection of $L_i$ with $\partial U_\dd$ changes.  At some size, however the intersection set is fixed -- there is only one singularity inside $U_\dd$.  More precisely, consider a continuous one-parameter family of $\epsilon$ neighbourhoods of the defect, $U_\dd(\epsilon)$. Then writing $\mathcal{B}_\dd(\epsilon) = \partial U_\dd(\epsilon)$ we can look at the intersection $\mathcal{B}_\dd(\epsilon) \cap L_i$. This defines a set of closed codimension-1 submanifolds $\Gamma_\dd(\epsilon)$ of $\mathcal{B}_\dd(\epsilon)$. There is then an $\epsilon_0$ such that $\Gamma_\dd(\epsilon)$ is isotopic to $\Gamma_\dd(\epsilon_0)$ for every $0< \epsilon \leq \epsilon_0$. We will use this invariant set to characterise the local structure of the defect in terms of the topology of $\mathcal{B}_\dd = \mathcal{B}_\dd(\epsilon_0)$ and the isotopy class of $\Gamma_\dd = \Gamma_\dd(\epsilon_0)$ (note that since $\Omega$ is either two or three dimensional, the Jordan-Sch\"{o}enflies  closed-curve theorem applies to $\Gamma_\dd$). 

\begin{figure}
\begin{center}
\includegraphics{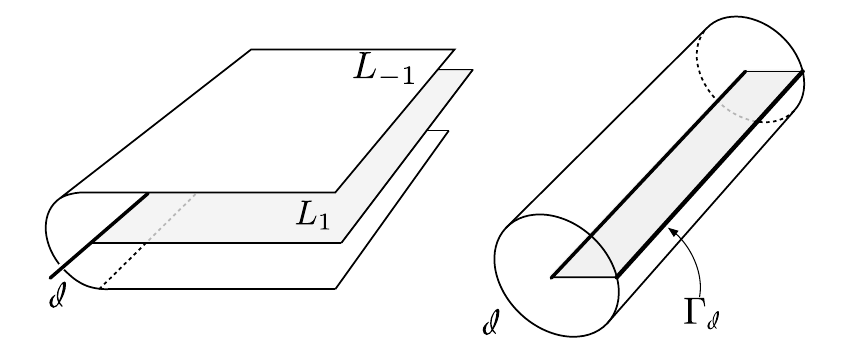}\\
\end{center}
\caption{\textit{Left:} Local layer structure for a $+1/2$ disclination line $\dd$ on $L_1$. \textit{Right:} Intersection, $\Gamma_\dd$, of $L_1$ with the boundary of the neighbourhood of the $+1/2$ defect line $\dd$. If the defect is a loop, $\Gamma_\dd\cong S^1$.}
\label{fig:ph_ex}
\end{figure}

We give two examples of this. First, consider a hyperbolic point defect, as found at the centre of focal conic domains. At the singularity, two layers meet at a point defect, forming a cone. The local structure of this cone is the zero set of the Morse-type singularity, $\phi = x^2+y^2-z^2$, with the point defect at the origin. $\Gamma_\dd$ is  the intersection of this cone with a small sphere and consists of two circles at height $z = \pm \sqrt{x^2+y^2}$. As a second example, consider a $+1/2$ disclination loop, illustrated in Figure \ref{fig:ph_ex}, in this case $\Gamma_\dd$ is a longitudinal circle on the toroidal neighbourhood of the defect loop.  This case also demonstrates that we can consider line defects in addition to point defects. 
\label{ex:poh}

We now define topological equivalence between defects. Two smectic textures are equivalent if there exists a one-parameter families of smectic configurations $\Psi_t$, such that no defect is created or removed, no defects coalesce, and the homotopy type of the defect sets themselves do not change.  We note that this does permit, for example, a point defect expanding into a line segment with two ends, but does not allow the transformation of a closed loop disclination into a point defect (or {\sl vice versa}), or the knot type of a disclination loop to change.

It follows from this that the isotopy class of the pair $(\mathcal{B}_\dd, \Gamma_\dd)$ cannot change under such a deformation. Any change in the isotopy class of $\Gamma_\dd$ would require an intersection between components of $\Gamma_\dd$, implying that an additional defect must either enter or leave $U_\dd (\epsilon)$. It follows that $(\mathcal{B}_\dd, \Gamma_\dd)$ constitutes a topological invariant of the defect. Since $(L_1, L_{-1})$ are sufficient to reconstruct the smectic phase field we find that, in the present context, they constitute a complete topological characterisation for smectic defects.

\subsection{Point Defects in Two Dimensions}
\begin{figure}[h!]
\begin{center}
\includegraphics{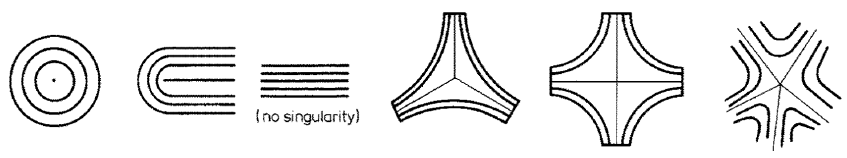}
\end{center}
\caption{Point defects in two dimensional smectics. From left to right they are given by the non-negative integer $m \in[0,5]$, with winding number $q=1-m/2$. Note that $m=2$ corresponds to no singularity. After Po\'{e}naru~\cite{poenaru81}.}
\label{fig:poenaru}
\end{figure}

In two dimensions we immediately recover Po\'{e}naru's classification result. In particular, let $\dd$ be a point defect, and $\gamma$ a circle around it. Then $\Gamma = L|_\gamma$ is a collection of $m \geq 0$ points. There is a unique embedding (up to permutation) of points on a circle so we find a correspondence between two dimensional smectic defects and non-negative integers. Each time one of these points is crossed while traversing $\gamma$, the phase field changes sign.  It follows that between two consecutive points that the director field rotates clockwise by $\pi$ in addition to the rotation angle between the two points.  One can immediately observe the winding number $q$ (defined as the degree of the map $N_\gamma:S^1 \to \mathbb{RP}^1$) of a two-dimensional disclination is
\begin{equation}
q = \frac{1}{2\pi} \left(2\pi - m\pi\right)=1-\frac{m}{2}
\end{equation}
so that $q \leq 1$. Figure \ref{fig:poenaru}, reproduced from \cite{poenaru81}, shows examples of two-dimensional point defects in smectics for $m \in [0,5]$. Note that the case $m=2$ does not correspond to a defect, {\sl per se}, but considering them as defects when necessary will enable a convenient description of defect combination rules in \S\ref{sec:rules}.

\subsection{Point Defects in Three Dimensions}
\begin{figure}
	\begin{center}
		\includegraphics[scale=0.4]{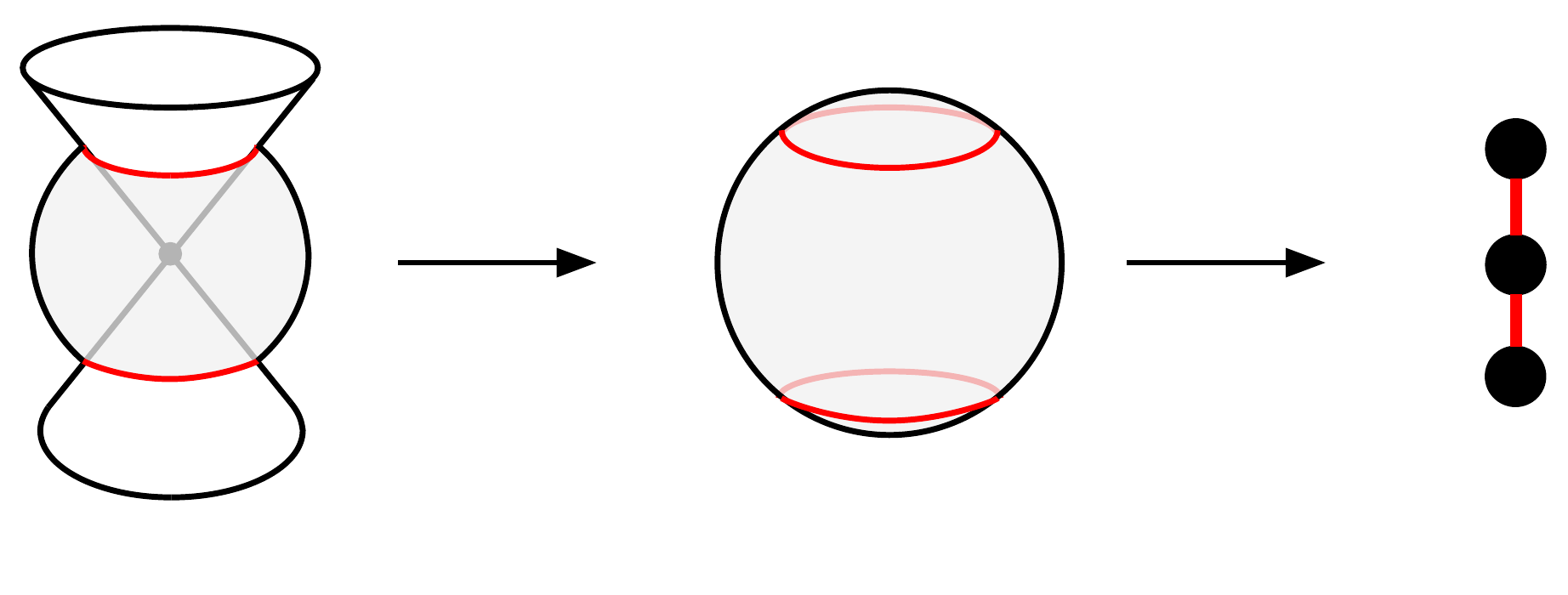}
	\end{center}
	\caption{The local structure of a point defect in a smectic can be described by a set $\Gamma$ of closed curves on a sphere (center, $\Gamma$ shown in red), or equivalently by a tree (right). In the case shown here, the point defect is described by the phase field $\phi = x^2+y^2-z^2$, and the resulting tree has three nodes.}
	\label{fig:pd_tree}
\end{figure}

\begin{figure}
\begin{center}
\includegraphics{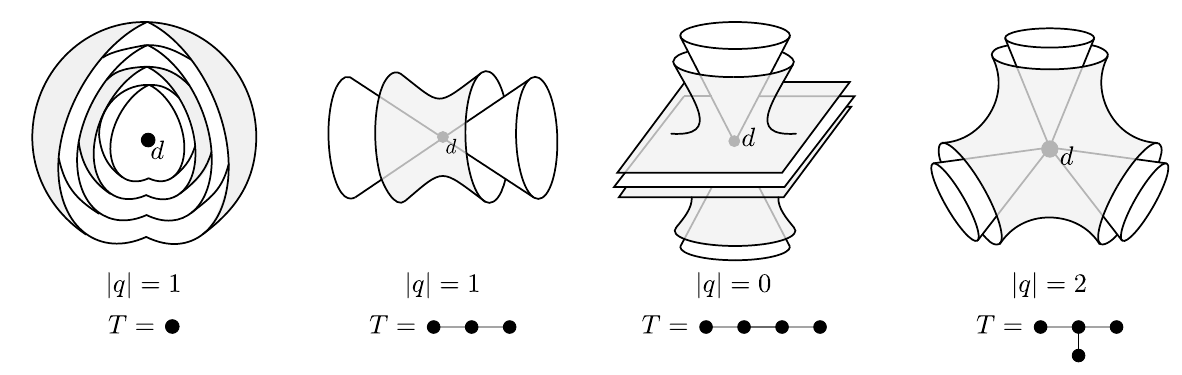}
\end{center}
\caption{Point defects in three dimensional smectics. The charge of the defect is shown by $|q|$, with $T$ the corresponding tree. Note that contrary to the two dimensional case, point defects in three dimensions are not uniquely identified by their charge.}
\label{fig:pd}
\end{figure}

In three dimensions, our results tell us that the classification of point defects is given by isotopy classes of non-intersecting closed curves on a sphere. There is a one-to-one correspondence between isotopy classes of closed curves on a sphere and trees, as illustrated in Figure \ref{fig:pd_tree}. In particular, let $\Gamma = \sqcup_i \Gamma_i$ be the curves on $S^2$ and $\mathcal{R} = \sqcup_j R_j = S^2 \setminus \Gamma$ their complement. Then draw a node for each component of $R_j$ and connect them with an edge if they are connected by some element in $\Gamma_i$. As $S^2$ is simply connected, the resulting graph is a tree. A set of examples is given in Figure \ref{fig:pd}. 

The charge of such a defect may be computed in terms of the tree representation by \eqref{eq:final_charge}. Let $\dd$ be a point defect in a three-dimensional smectic with tree representation $T$. Then, up to sign the charge, $q$, of the point defect is given by
\begin{equation}
| q | =\left | \sum_{v \in \mathcal{V}(T)} s(v) \left (1-\frac{d(v)}{2} \right) \right |
\label{eq:final_charge}
\end{equation}
where $\mathcal{V}$ is the vertex set of $T$, $d(\cdot)$ gives the degree of a vertex, and $s(\cdot)$ gives the sign of a vertex in an alternating assignment of sign to each vertex. To show \eqref{eq:final_charge} we assume without loss of generality that the defect is described by a function $\phi : (\mathbb{R}^3,0) \to (\mathbb{R},0)$ with an isolated critical point at the origin and a specified set $\phi^{-1}(0)$ along with an equivalence relation $\phi \sim - \phi$. The charge of the point defect is then determined up to sign by the degree of the map ${\bf n} : S_\epsilon^2 \to S^2$ given by the director field (normalised gradient of $\phi$), where $S_\epsilon^2$ is the $\epsilon$-sphere around the origin. 
Now, let $H_{{\bf n},\epsilon}$ be the 2-plane bundle over $S^2_\epsilon$ corresponding to vectors orthogonal to $\bf n$. The Euler class of this bundle is just the signed area$/2\pi$ swept out by $\bf n$ on the unit sphere and generalizes the Euler characteristic, the Euler class of a manifold's tangent bundle. We have: 
\begin{equation} e(H_{{\bf n}, \epsilon}) = 2 q
\label{eq:euler}
\end{equation}
where $e(\cdot)$ is the Euler class. Now let $ \bf{r}$ be the radial vector in $\mathbb{R}^3$, and define the vector field ${\bf v}$ on $S^2_\epsilon$ by
\begin{equation}{\bf v} ={\bf r} \times {\bf n}
\end{equation}
which is a section of $H_{{\bf n}, \epsilon}$ and $TS^2_\epsilon$. Then from \eqref{eq:euler} and the definition of the Euler class we have
$$ q = \frac{1}{2}\sum_{i} \textrm{Sgn}_i\textrm{Ind}(x_i) $$
where $x_i$ are the zeros of ${\bf v}$,  $\textrm{Ind}(x_i)$ are their windings, computed using the orientation of $S^2_\epsilon$ and $\textrm{Sgn}_i= \textrm{Sgn}({\bf r} \cdot {\bf n})|_{x_i}$ determines the relative orientation of $S^2_\epsilon$ and $H_{N,\epsilon}$ at the zeros of ${\bf v}$. Now ${\bf r} \cdot {\bf n} \approx k \phi(x)$ for $\| x \| $ small, where $k$ is the order of the lowest non-zero gradient, so that for $\epsilon$ sufficiently small $\textrm{Sgn}({\bf r} \cdot {\bf n})$ is constant in each region, $R_j$, and must alternate between neighbouring regions. Finally, note that ${\bf v}$ is tangent to $\Gamma$, so that the sum of windings $\textrm{Ind}(x_i)$ in a region is just equal to the Euler characteristic of the region, $\chi(R_j) = 2-b_j$, where $b$ is the number of boundary components. It follows that
\begin{equation}
q = \frac{1}{2} \sum_j \textrm{Sgn}(R_j) \chi (R_j),
\label{eq:end}
\end{equation}
where $\textrm{Sgn}(R_j)$ is the sign of $\phi$ in $R_j$. Allowing for the global $\mathbb{Z}_2$ action and passing to the tree representation establishes the equivalence of \eqref{eq:end} and \eqref{eq:final_charge}.
Under the symmetry $\phi \to - \phi$, $q \to -q$, so that the charge of free homotopy classes of point defects is determined only up to sign. As we will see, when adding defects together, one must consider based homotopies so that the relative charge between defects is well-defined.

The classification of point defects in three dimensional smectics is fundamentally different from the two dimensional case. In two dimensions the winding numbers of defects are {\it a priori} integers, elements of $\pi_1(\mathbb{RP}^1)= \mathbb{Z}$. The constraints enforced by the smectic mean that not all integers are realisable in a smectic, but the charge is still a unique identifier for the defect topology. In three dimensions the opposite is true. The charge of the point defect {\it a priori} lives in the set $\pi_2(\mathbb{RP}^2)/ (q \sim -q)$, and all elements of the set are realisable in a smectic. In fact, as we have seen, the charge is no longer a unique identifier of topological type for point defects in three-dimensional smectics, indeed there are an infinite number of topologically distinct defects for each possible charge. 

The origin of this difference between two and three dimensions can already be seen in the structure of generic (Morse) type singularities, and ultimately comes from the fact that spatial inversion ${\bf x} \to - {\bf x}$ reverses orientation in three dimensions but not two. A Morse singularity with Morse index $m$ has charge $q = (-1)^m$ and under the smectic symmetry of $\phi \to - \phi$ the Morse index transforms to $d-m$ and describes an equivalent smectic configuration. If $d$ is even this transformation leaves the charge invariant, and in odd dimensions it changes sign. Specifically, in two-dimensions, the $m=2$ and $m=0$ defects have charge $+1$ and cannot be added together -- they correspond to top and bottom handles and cannot merge with themselves or with each other.  Only the $m=1$ saddles can merge leading to arbitrary negative-charged defects.  In three-dimensions, there are two saddles, $m=1$ and $m=2$, with opposite point charges.  These can be joined with themselves to form point charges of arbitrary charge.  However, because there is more than one way to assemble them,  the charge of a smectic point defect cannot capture its entire topology.

\subsection{Line Defects in Three Dimensions}

As extended objects, line defects have a richer structure than point defects and the corresponding classification is commensurately more complex. Firstly, a closed loop line defect may be tied into a knot, and so the knot type becomes a topological invariant of the defect, per our definition. We will not dwell on the topic of knotted disclinations in smectics, noting only that there are subtle global constraints that are not yet fully understood. In particular, one may think of a smectic configuration with a single knotted disclination line. If the knot is not fibered, then there must be additional defects in the system, or a more complex structure along the defect itself~\cite{aharoni17,dennis17,kamien16} such as monopoles, for example. 

\subsubsection{Global Structure of Line Defects}

\begin{figure}
\begin{center}
\includegraphics{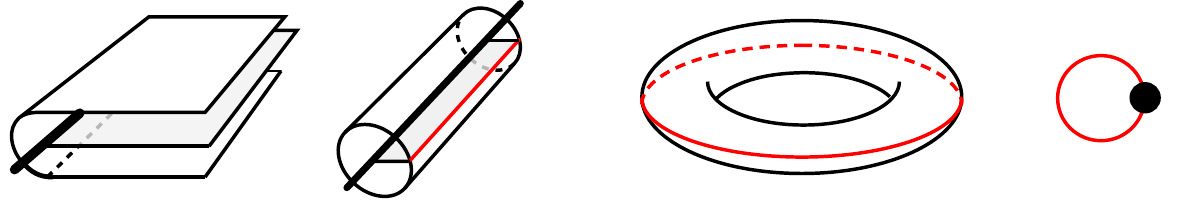}
\end{center}
\caption{Classifying line defects, $\Gamma$ defines a set of closed curves (shown in red) on a torus. These define a graph with up to one cycle, via a procedure identical to the case of point defects. An additional integer invariant coming from the $SL(2,\mathbb{Z})$ action on $U_\dd(\epsilon)$ specifies the structure.}
\label{fig:lines_tree}
\end{figure}

Neglecting the knot type of the defect, we can examine the defect structure for a line defect analogously to our study of point defects. $\mathcal{B}_\dd$ is now a torus, and so we are led to consider isotopy classes of nonintersecting closed curves on the torus, as shown in Figure \ref{fig:lines_tree}. These have a partial graph-theoretic representation as graphs with up to one cycle (Betti numbers $b_0 = 1$, $b_1 \leq 1$). There is an additional invariant coming from the $SL(2,\mathbb{Z})$ action on $U_\dd(\epsilon)$, which quantifies the twisting of the layers along the longitude of the defect line and is well defined as we take $\Omega \subseteq \mathbb{R}^3$. Specifically, if the graph contains a cycle of length $l$, then the additional invariant is best described by a rational number $p\in \mathbb{Q}$ which is the total rotation (in units of $2\pi$) of the layers along the cycle. The defect line in this case is a $(1-m/2)$ disclination line, with $m=l\cdot\mathrm{denominator}(p)$ the number of layers attached to $\dd$. A defect loop with $p \neq 0$ must link another defect (a dislocation). Such situations can arise when an edge dislocation links a screw dislocation, for example~\cite{kamien16,blanc04}.

\begin{figure}
\begin{center}
\includegraphics{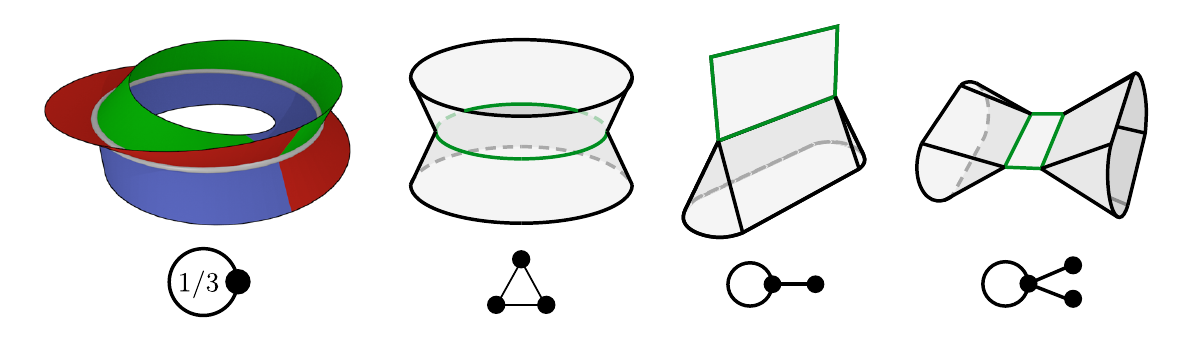}
\end{center}
\caption{Examples of layer structure around smectic line defects. Left to Right: a $-1/2$ disclination line with a $2 \pi/3$ twist; a $-1/2$ disclination line without twist, this can be viewed as an expanded focal conic core; a disclination line with a mixture of $+1/2$ and $-1/2$ profiles; a disclination line with two $-1/2$ regions and two $+1/2$ regions, this line is also an expanded focal conic core.}
\label{fig:line_defects_gallery}
\end{figure}

Figure~\ref{fig:line_defects_gallery} shows several examples of defect line structures and their corresponding graph theoretic representations, which now include a single $l$-node loop with the additional invariant $p$ indicated if it is non-zero, as in the leftmost example in Figure~\ref{fig:line_defects_gallery}. In the remaining examples there is a single layer spanning the defect loop, so that the twisting invariant vanish. In such cases, one can consider the result of shrinking this central layer, recovering a point defect. This can be achieved simply by removing the edge in the graph corresponding to the spanning layer. In the cases shown in Figure~\ref{fig:line_defects_gallery} this results in either a hyperbolic point defect (left and right) or a simple null-defect. The hyperbolic point defects can be considered as focal conic domain cores, and so the line defects on the left and right may be thought of as expanded versions of these focal conic cores

\subsubsection{The Local Structure of Line Defects}

\begin{figure}
\begin{center}
\includegraphics{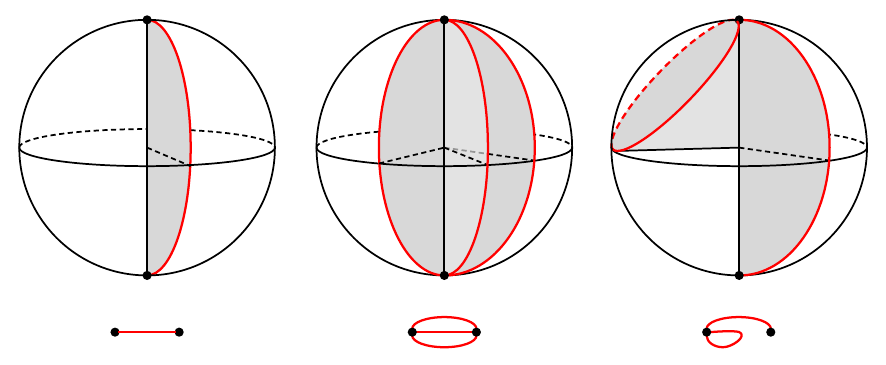}
\end{center}
\caption{Local structure of line defects in three dimensions. Left to right: local structure of a $+1/2$ disclination line; local structure of a $-1/2$ disclination line; local structure showing the transition from a $+1/2$ to $-1/2$ disclination, which must occur at a discrete point in a smectic. Below each diagram is a schematic representation of the structure.}
\label{fig:local_structure_lines}
\end{figure}

Alternatively, we may examine the local structure of line defects. Consider a point $x$ on a line defect, and a small sphere $S^2_\epsilon$ centred at $x$. Then, as in the previous cases, there will be an $\epsilon_0$, such that for $0<\epsilon \leq \epsilon_0$, the topology of the intersection of $L_i$ with $S^2_\epsilon$ will not change, and so gives a description of the local structure of the disclination line at $x$. This case differs from the previous ones in that the defect line itself intersects $S^2_\epsilon$. As a consequence of this, the local structure of point on a line defect is given by isotopy classes of closed curves on a sphere with two marked points, corresponding to the intersection of the defect line with the surrounding sphere, where closed curves, arising from cone-like branches of the smectic layers, may end.  This  Figure \ref{fig:local_structure_lines} shows examples of this structure for the local structure of a $+1/2$ disclination line, a $-1/2$ disclination line and the local structure of a transition between $+1/2$ and $-1/2$ profiles.

In this case, we have no simple graph-theoretic representation of the local defect structure, in contrast to the case of point defects and (partially) line defects, though the local structures may be represented as diagrams of curves in the plane with two marked points on which curves may end (as shown in Figure \ref{fig:local_structure_lines}). If $\mathcal{P}$ represents the set of isotopy classes of such diagrams, then a parameterisation of a closed loop line defect gives a map $f:S^1 \to \mathcal{P}$. Since $\mathcal{P}$ is a discrete space, $f$ is locally constant, and must change at discrete points. Consequently we find that in three dimensional smectics a line defect comes with a set of special points at which the local structure of the defect changes. An immediate physical consequence of this comes when considering profile transitions~\cite{aharoni17}. Consider a line defect in a nematic and the restriction of the director field to a small disc perpendicular to the line.   If the director is tangent to this disc then we could assign a geometric winding  $m\in \mathbb{Z}+1/2$  However, since $\pi_1(\mathbb{RP}^2) = \mathbb{Z}_2$ it is possible to have another disc along the line with any other winding $m'\in \mathbb{Z}+1/2$.  Between these discs the texture smoothly deforms via ``escape into the third dimension.'' In the smectic case, however, this is not possible. The transition from, for instance, a local $+1/2$ profile (Figure \ref{fig:local_structure_lines}, left) to a $-1/2$ profile (Figure \ref{fig:local_structure_lines}, center) must occur via a monopole like structure (Figure \ref{fig:local_structure_lines}, right), a discrete object sitting on the disclination line itself that, in principle, should be experimentally observable. 

\section{Combination Rules for Point Defects}\label{sec:rules}

It is unsurprising that the intricate classification of point defects in smectics precludes a simple rule for their combination. As a motivating example it is instructive to consider a simplified case: the addition of two $+1$ disclinations in a two-dimensional system. For this example we relax our assumption that defects lie on density extrema and consider the smectic phase field as a smooth real-valued function. It is a consequence of the mountain pass theorem that any two $+1$ disclinations must entrain a $-1$ (the mountain pass) and so bringing two $+1$ disclinations together may only be done in the presence of the $-1$, resulting in a single $+1$ disclination.

Reemploying the constraint that disclinations lie on density extrema, a smooth deformation of the texture considered in the previous paragraph is no longer possible. As the defects are brought closer together the gradients of $\phi$ must diverge, with $\phi$ becoming discontinuous in the limiting case. A finite compression modulus implies a large energy cost for this process, with it ultimately being censored by the smectic melting.

The broken translational symmetry of the smectic lies at the heart of this difficulty, combining defects along paths transverse to layers forces layer compression into the system. This can be evaded by combining defects along level sets of the smectic density, where there is no obstruction to defect addition. This distinction underlies an essential dichotomy between transverse addition, where defects are combined transverse to layers, and tangential, where defects are combined along layers.

Tangential addition of disclinations is relatively simple and can be completely characterised by a connected sum operation on the local defect structures. By contrast, transverse addition is considerably more complex, the local addition operation does not na\"{i}vely conserve charge, and we find that additional defects must be created. The nature of these additional defects depends sensitively on the path along which the defects are combined, conferring an intrinsic path dependence on transverse defect addition. This is similar in form to the case of translationally invariant systems whose groundstate manifold possesses a non-Abelian fundamental group~\cite{poenaru77}, but its origins are distinct. More importantly, even if we start with a texture that is globally orientable, defect combination can require the presence of half-integer disclinations.  We can thus define, {\sl en passant}, the ``sterile smectic,'' one which only allows tangential addition.

\subsection{Defining Addition}

We define the addition of defects in terms of their local structure. In \S\ref{sec:classification}, we represented a defect, $\dd_a$, as a pair $(\mathcal{B}_a, \Gamma_a)$, where $\mathcal{B}_a$ was the boundary of a sufficiently small neighbourhood of the defect and $\Gamma_a = \sqcup_i \Gamma_{ai}$ a closed codimension-one submanifold of $\mathcal{B}_a$ representing the intersection of the layers at the defect with $\mathcal{B}_a$. In particular $\Gamma_a$ defines a decomposition of $\mathcal{B}_a$ into regions
\begin{equation}
\mathcal{B}_a \setminus \Gamma_a = \sqcup_j R_{aj}.
\end{equation}

\begin{figure}
\begin{center}
\includegraphics{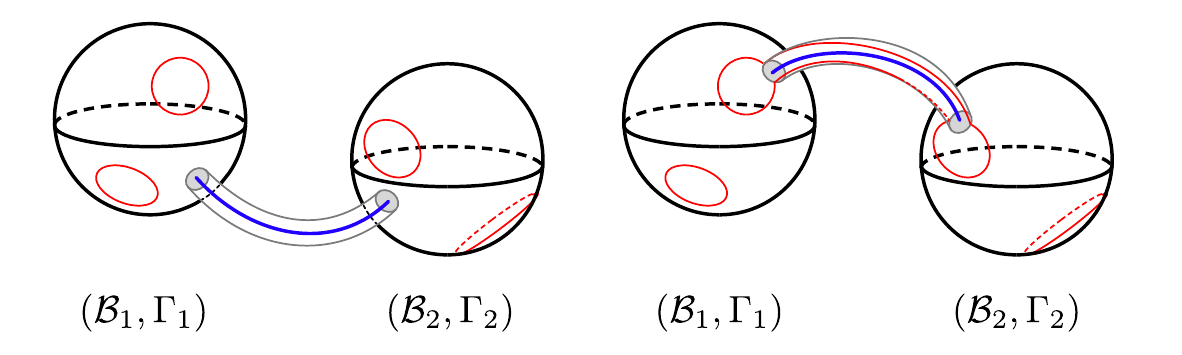}
\end{center}
\caption{{\em Left:} Connected sum illustrating transverse addition of local defect structures. {\em Right:} Connected sum illustrating tangential addition of local defect structures, note that this must descend to a consistent connect sum operation on the sets $\Gamma_1$ and $\Gamma_2$. }
\label{fig:con_sum}
\end{figure}

Defect addition is naturally represented via connected sums of the pairs $(\mathcal{B}_a, \Gamma_a)$ for each defect. If $\dd_1$ and $\dd_2$ are two defects, then combining them along the path $c$ is realised as a connected sum of $(\mathcal{B}_1,\Gamma_1)$ and $(\mathcal{B}_2,\Gamma_2)$ around the points $x = c \cap \mathcal{B}_1$ and $y = c \cap \mathcal{B}_2$, as shown in Figure \ref{fig:con_sum}. There are several restrictions. Firstly, both $\dd_1$ and $\dd_2$ must either both be density minima or both be density maxima, otherwise their combination would result in a phase singularity, which we forbid. Secondly, the path $c$ may intersect components of $L$ that are not locally connected to either $\dd_1$ or $\dd_2$. These intersections will not be considered in the defect addition itself, their effect is studied in section \ref{sec:homotopy}. Finally, the connected sum operation must respect the $\Gamma_a$. This separates the possible additions into two types: those in which both $x$ and $y$ are elements of $\Gamma_a$, corresponding to tangential addition, and those for which both $x$ and $y$ are not, corresponding to transverse addition. 

We will denote $\dd_1 \pitchfork_{j,j'} \dd_2$ the transverse addition of $\dd_1$ and $\dd_2$, realised as the connected sum of $(\mathcal{B}_1,\Gamma_1)$ with $(\mathcal{B}_2,\Gamma_2)$ at points $x \in R_{1j} \subset \mathcal{B}_1$ and $y \in R_{2j'} \subset \mathcal{B}_2$. Tangential addition needs a little more care to define properly, the connected sum must respect the sets $\Gamma_1$ and $\Gamma_2$. As a consequence, in three dimensions there are two distinct additions for each choice of $x \in \Gamma_1$ and $y \in \Gamma_2$, related to the two possible choices of alternating sign in the tree representation. These may be distinguished by giving each component of $\Gamma_{a}$ an orientation, whereupon the two addition operations may be seen to either match or not the orientations of the $\Gamma_{a}$. With this in mind, we write ${\dd}_1 \parallel_{{i,i'}, \pm} {\dd}_2$ to denote the tangential addition of ${\dd}_1$ and ${\dd}_2$ along layer components $\Gamma_{1i}$ and $\Gamma_{2i'}$, with the $\pm$ sign denoting and orientation preserving or reversing addition in three dimensions. If $\Gamma_{ai}$ or $R_{aj}$ are unspecified, we will simply use $\dd_1 \parallel \dd_2$ or $\dd_1 \pitchfork \dd_2$. Note that if the regions in both pairs $(\mathcal{B}_a,\Gamma_a)$ are given alternating signs, then this may be used to define orientations for the $\Gamma_a$. The orientation preserving transverse addition then allows for a consistent formula for charge addition.

\subsection{Two Dimensions}

In two dimensions, the result of defect addition is easy to compute. Point defects are classified by non-negative integers. The addition of two point defects $\dd_1$ and $\dd_2$ with $m_1$ and $m_2$ prongs respectively results in a point defect with $m_1+m_2$ prongs if the defects are combined transversely or $m_1+m_2-2$ prongs if they are combined tangentially. This implies the following relations:
\begin{equation}
q(\dd_{1} \parallel \dd_2) = q(\dd_1)+q(\dd_2)
\end{equation}
\begin{equation}
q(\dd_{1} \pitchfork \dd_2) = q(\dd_1)+q(\dd_2)-1.
\label{eq:trans2D}
\end{equation}
Note that transverse addition does not conserve charge {\sl locally}, and therefore additional defects must be created, a description is given in \S\ref{sec:homotopy}. One interesting corollary of \eqref{eq:trans2D} is that the combination of any defect with a $+1$ defect, which can only combine transversely, acts as the identity; the original defect is restored and a new $+1$ is created.

\subsection{Three Dimensions}
In three dimensions the situation is more complex. The charge of a point defect is dependent on a choice of alternating sign for each region, and in order to define a consistent addition process, the signs of the regions participating in the connected sum must match. This is equivalent to defining a basepoint for point defect addition in nematics. 

\begin{figure}
\begin{center}
\includegraphics[scale=0.8]{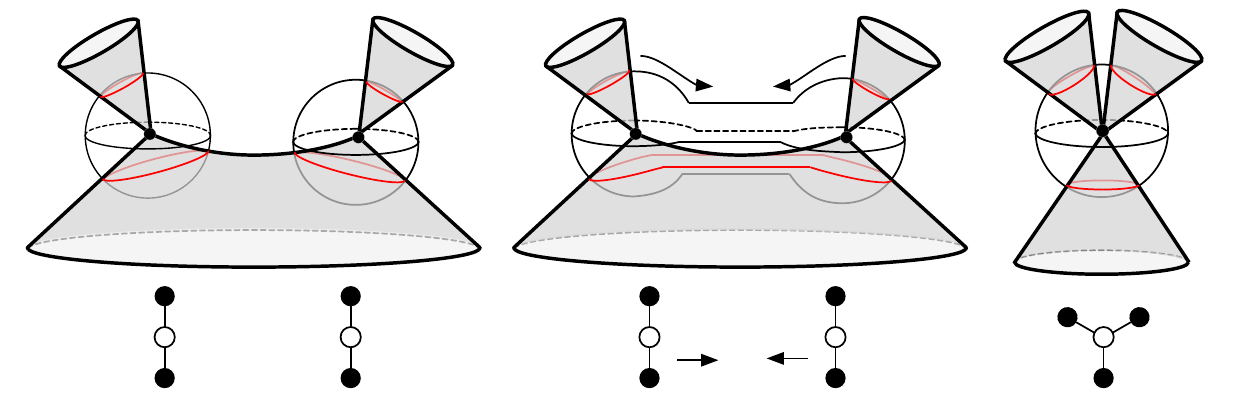}
\end{center}
\caption{Tangential addition of point defects in 3D smectics realised as edge union of graphs.}
\label{fig:3d_comb}
\end{figure}

Given a signed pair $(\mathcal{B}_a, \Gamma_a)$, we can compute the charge $q(\dd_a)$ using \eqref{eq:end} without the modulus operation. We are then in a position to define the addition. We will first consider tangential addition of point defects illustrated schematically in Figure~\ref{fig:3d_comb}. It is sufficient to consider the effect on the tree representation, and it is easy to show that tangential addition of point defects is realised as edge-wise union in the tree representation. The connected sum of $\mathcal{B}_1$ and $\mathcal{B}_2$ descends to a connected sum on two components of $\Gamma_1$ and $\Gamma_2$, resulting in an identification of edges in the tree representation. Note that the sign ambiguity in addition of 3D point defects corresponds to the two possible ways one can form an edge-wise union of two graphs with two edges. Specifying the signs of the regions of $(\mathcal{B}_a, \Gamma_a)$ removes this freedom, leaving only one valid addition. In this case, it is easy to check that the charge of the resulting addition is 
\begin{equation}
|q(\dd_1 \parallel \dd_2)| = |q(\dd_1)+ q(\dd_2)|.
\end{equation}

\begin{figure}
\begin{center}
\includegraphics{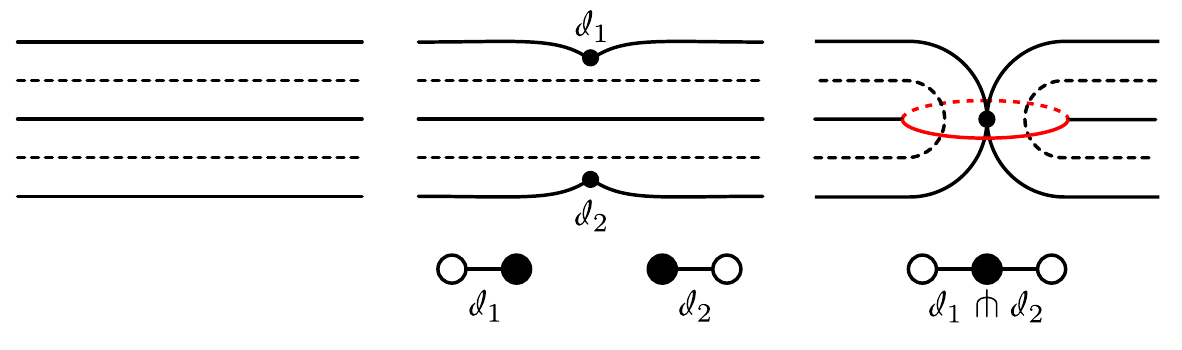}\\
\end{center}
\caption{An example of transverse addition of point defects in 3D smectics creating a focal conic domain. $L_1$ is shown by the thick black lines, $L_{-1}$ by the dotted lines. From left to right: a smectic groundstate; homotopically trivial defects $\dd_1$ and $\dd_2$ are created, they are cusps in the layers, and can be represented as two node trees; $\dd_1$ and $\dd_2$ are combined vertically, creating a homotopically non-trivial hyperbolic defect, with absolute charge $1$, realised as the node union of the trees. The process must create an additional defect, as discussed in \ref{sec:homotopy} this must be a line defect, demonstrating that the point defects are not closed under addition in smectics.}
\label{fig:3d_combt}
\end{figure}

By a similar token, and illustrated in Figure~\ref{fig:3d_combt}, transverse addition of point defects in three dimensions acts as node-wise union in the tree representation, as the connected sum identifies the regions $R_{1j}$ and $R_{2j'}$, which are represented as nodes. Again, a consistent addition requires the signs of $R_{1j}$ and $R_{2j'}$ match. The resulting charge of the point defect depends on the two nodes identified, and may be written as
\begin{equation}
|q(\dd_{1} \pitchfork \dd_2)| = | q(\dd_1) + q(\dd_2)- 1|
\label{eq:comb}
\end{equation}

\subsection{Homotopy Theoretic Aspects of Transverse Addition}
\label{sec:homotopy}

The charge formulae for transverse addition of disclinations do not conserve charge, with extra defects of charge $+1$ required. Since the defect combination process is inherently singular, there is no way to unambiguously specify the additional defects created by this process. However, several of their properties are determined by topological invariants associated to the path between the two defects that are being combined. We note also that one may develop a scheme of this type for general translationally ordered systems~\cite{prep}. 

The path-dependence of transverse addition comes through an element of $\pi_1(E_d)$ that one may associate with transverse paths between disclinations, with $E_d$ the groundstate manifold of the smectic in $d$ dimensions. On a heuristic level, this group element can be seen to measure the number of layers and winding of the director field between two smectic disclinations. It is this group element which controls the additional defects created. For reference it is useful at this point to state some of the topological properties of $E_d$. For $d=2$, $E_d$ is the Klein bottle, so we have
\begin{equation}
\pi_1(E_2) \cong \mathbb{Z} \rtimes \mathbb{Z}.
\end{equation}
in three dimensions, $E_3$ is a twisted circle bundle over $\mathbb{RP}^2$, the fundamental group can be computed directly and shown to be the infinite dihedral group~\cite{kleman78},
\begin{equation}
\pi_1(E_3) \cong D_\infty \cong \mathbb{Z} \rtimes \mathbb{Z}_2.
\end{equation}
The fibre bundle structure of $E_d$ gives a natural projection map
\begin{equation}
\rho: E_d \to \mathbb{RP}^{d-1}
\label{eq:proj0}
\end{equation}	
and as a result, by abuse of notation, the maps
\begin{equation}
\rho: \pi_n(E_d) \to \pi_n(\mathbb{RP}^{d-1})
\label{eq:proj}
\end{equation}
which simply forget about the phase information, only record the topology of the director field. 

\begin{figure}
\begin{center}
\includegraphics{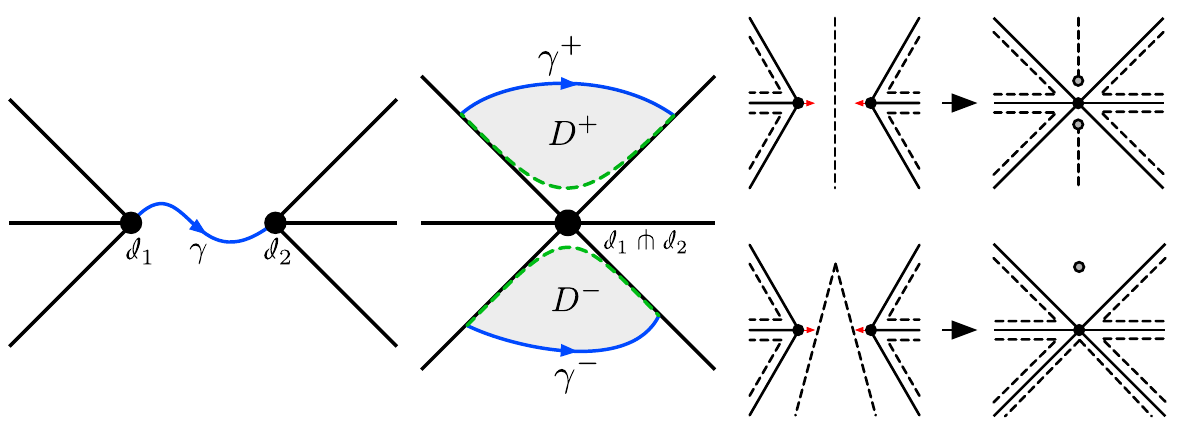}
\end{center}
\caption{Combining defects $\dd_1$ and $\dd_2$ in two dimensions. {\it Left \& Centre}: $\gamma$ is a transverse path between two defects, $\dd_1$ and $\dd_2$. Associated to $\gamma$ is an element $g_\gamma \in \pi_1(E_d)$. After the transverse addition of $\dd_1$ and $\dd_2$, additional defects are forced to be created in the regions $D^+$ and $D^-$. The element of $\pi_1$ associated to $\partial D^{\pm}$ dictates the additional defects created in each region via \eqref{eq:extra}. A similar picture can be drawn in three dimensions, only now $D^+$ and $D^-$ are connected.  {\it Right, Top}: Transverse combination of two $-1/2$ defects in two dimensions. In this case $\rho(g_\gamma) = 0 \in \mathbb{Z}/2$, so that both $D^+$ and $D^-$ contain a $+1/2$ defect.  {\it Right, Bottom}: Transverse combination of two $-1/2$ defects in two dimensions. In this case $\rho(g_\gamma) = -1/2 \in \mathbb{Z}/2$, so that $D^+$ contains a single $+1$ defect, and $D^-$ is a defect free region.  }
\label{fig:2d_extra}
\end{figure}

We now show how to associate an element of $\pi_1(E_d)$ to transverse paths between layers. We will define a transverse path $\gamma:[0,1] \to \Omega$  between point defects $\dd_1,\dd_2\in L_{\pm 1}$ that does not intersect other defects. The sign of $L$ must be the same for $\dd_1$ and $\dd_2$ for their combination to not generate a phase singularity, which we forbid. $\gamma$ is transverse to $L$ at its endpoints, so that
\begin{equation}
\lim_{t \to 0,1} {\partial_t \gamma}\cdot{\bf n} \ne 0
\label{eq:trans_condition}
\end{equation}
where ${\partial_t \gamma}$ is the tangent to $\gamma(t)$, and ${\bf n}$ is the director field at $\gamma(t)\in\Omega$.

We can define a map $f:(0,1)\to E_d$ by composing $\gamma$ with the map $\Omega \setminus \mathcal{D} \to E_d$ given by the smectic order, and complete $f$ continuously at its endpoints to obtain $f:[0,1]\to E_d$. We now wish to complete $f$ to a closed loop in $E_d$ by connecting $f(1)$ back to $f(0)$ in a canonical way, {\it i.e.}, to ``compare'' the smectic structure along $\gamma$ to a structure which is in some sense trivial. This can be achieved thanks to the transversality condition \eqref{eq:trans_condition}; neither $\rho\circ f(0)$ nor $\rho\circ f(1)$ are in $SN\gamma$, the unit normal bundle of $\gamma$ in $\Omega$. Roughly speaking, we can arrange the two defects so that $\gamma$ is the straight line connecting them and is perpendicular to both. We can use this to define a common base point in $\mathbb{RP}^{d-1}$ for the director field on the two defects. More precisely, since $PT_\gamma \Omega \setminus SN \gamma \cong (0,1) \times  D^{d-1}$ is simply connected, where $D^{d-1}$ is a disc and $PT_\gamma \Omega \cong \mathbb{RP}^{d-1}$ is the projectivised tangent space of $\Omega$ restricted to the interior of $\gamma$, there is a unique continuous path (up to homotopy) $\nu:[0,1]\to\mathbb{RP}^{d-1}$ such that $\nu(0)=\rho\circ f(0)$, $\nu(1)=\rho\circ f(1)$ and ${\partial_t \gamma}\cdot\nu \ne 0$ for every $t$. Namely, there is a topologically unique transport along $\gamma$ of the director near $\dd_1$ to the director near $\dd_2$ which extends the endpoint transversality to all of $\gamma$. We now define $\tilde{\nu}:[0,1]\to E_d$ to be $\tilde{\nu}(t)=(\nu(t),\phi_0)$, where the constant $\phi_0$ is the phase at both $\dd_1$ and $\dd_2$ (either density minimum or maximum). The composed path $f\tilde{\nu}^{-1}:S^1 \to E_d$ is an element $g_\gamma \in \pi_1(E_d)$ which does not depend on any of the choices made above. 

Of physical interest are the deformations of $\gamma$ for which $g_\gamma$ is invariant. Observe that the smectic order must be defined in the interior of $\gamma$, so that $\gamma$ may not pass through defects. Conversely, deforming $\gamma$ so that it passes through a defect will change $g_\gamma$ in much the same way as translationally invariant defects with non-Abelian fundamental group~\cite{poenaru77}. The second deformation of $\gamma$ which is of physical interest is associated to the endpoints of $\gamma$. $g_\gamma$ can only be defined if $\gamma$ satisfies the transversality condition \eqref{eq:trans_condition}, it follows that deformations of $\gamma$ that pass through configurations that break this condition may change $g_\gamma$. An important example of a deformation that doesn't break this condition is to move the endpoints of $\gamma$ along layers. One can also deform $\gamma$ while keeping its endpoints fixed, as long as the tangent ${\partial_t \gamma}$ is not tangential to layers at its endpoints. It is here, with $g_\gamma$ associated to the direction along which $\gamma$ approaches the defects, that the phenomena of path-dependence becomes truly distinct from the non-Abelian case, in that it is intrinsic to broken translational symmetry.

\begin{figure}
\begin{center}
\includegraphics{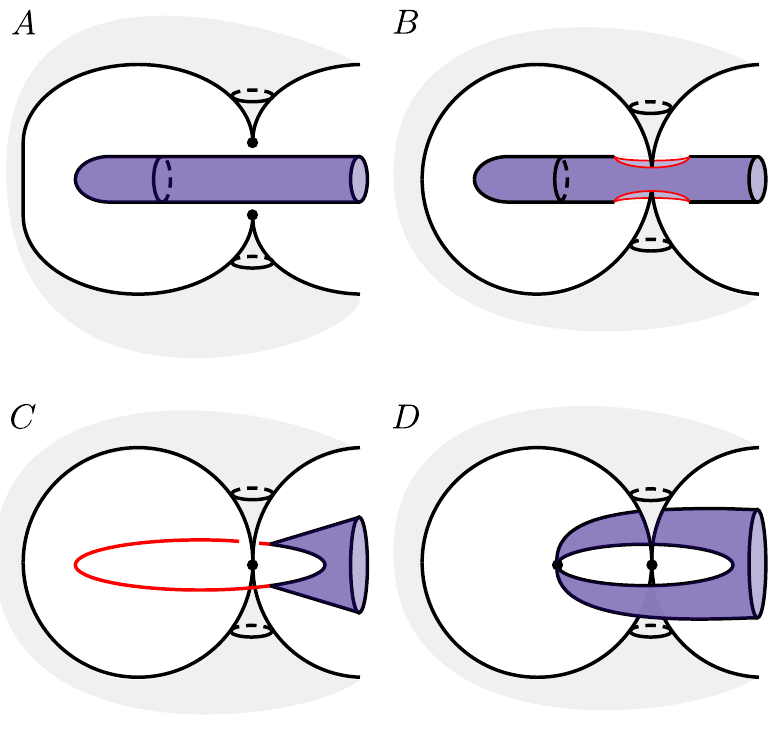}\\
\end{center}
\caption{Transverse defect combinations with $\rho(g_\gamma) \neq 0$. Combining two homotopically trivial defects (two node trees), c.f. Figure \ref{fig:3d_combt}. A: two homotopically trivial defect on the same layer, $L_1$ (gray), $L_{-1}$ is shown in violet. B: the two defects are combined transversely. Since $\rho(g_\gamma) = 1 \in \mathbb{Z}_2$, this requires the creation of additional defects with total hedgehog charge $|1|$, without an essential non-orientable disclination line. In this case, this is achieved by the creation of two defect lines, shown in red. C: the two defect lines are combined to form a $+1$ defect arc. D: the arc is contracted to form a single point defect.}
\label{fig:cybts}
\end{figure}

Given $g_\gamma$, we can determine the additional defects created by transverse addition. In this paper we will discuss the two and three dimensional cases, a more general formulation will be the topic of future work~\cite{prep}. For the two dimensional case, consider Figure \ref{fig:2d_extra}, where two defects, $\dd_1$ and $\dd_2$, are combined along a path $\gamma$, then let $g_\gamma \in \pi_1(E_2)$ be the group element associated to $\gamma$ as defined above. After the defect addition, the domain between $\dd_1$ and $\dd_2$ is split into two regions, and we may assume that any extra defects lie in compact regions $D^+$ and $D^-$ above and below the defect, so that far away from it the texture is unchanged. By integrating around either region $D^{\pm}$, the total disclination charge inside it is readily shown to be
\begin{equation}
q(D^\pm) = \frac{1}{2} \mp \rho(g_\gamma),
\label{eq:extra}
\end{equation}
where $\rho: \pi_1(E_2) \to \pi_1(\mathbb{RP}^1) = \mathbb{Z}/2$ is the projection map in \eqref{eq:proj}. There is an exception to this case. If either $\dd_1$ or $\dd_2$ is a $+1$ disclination, then $D^+$ and $D^-$ may not be defined. In this case, there is simply an additional $+1$ disclination created. Figure \ref{fig:2d_extra} shows examples of both a case where $\rho(g_\gamma)=0$, so that two $+1/2$ defects are created, and a case where $\rho(g_\gamma) \neq 0$ and only a single $+1$ defect is formed on one side.

In three dimensions, $\rho(g_\gamma) \in \pi_1(\mathbb{RP}^2) \cong \mathbb{Z}_2$, and the corresponding regions $D^\pm$ are connected, forming a solid torus. By analogous reasoning to the two dimensional case, the $\mathbb{Z}_2$ invariant determines whether a line defect is created via the addition operation or not. An example of this operation is shown in Figure \ref{fig:3d_combt}, where we see an additional line defect created. Figure \ref{fig:cybts} shows an example of defect combinations where the line defect is not topologically required. Finally, we note that we have not made use of all the information in $g_\gamma$, which additionally records the number of smectic layers between defects, and may be associated to Burgers vector-like quantities arising from defect combinations~\cite{prep}.

We would like to acknowledge conversations with B.G. Chen and R.A. Mosna. This work was partially supported by the NSF through Grant DMR-1262047 as well as a Simons Investigator award from the Simons Foundation
to R.D.K.


\begin{thebibliography}{99}
\bibitem{deGennesLC} P.G. de Gennes, {\it The Physics of Liquid Crystals}, 2nd ed.\ (Clarendon Press; Oxford, United Kingdom; 1995).
\bibitem{deGennes72} P.G. de Gennes, Solid State Commun. {\bf 10}, 753 (1972).
\bibitem{chen09} B.G. Chen, G.P. Alexander and R.D. Kamien, Proc. Natl. Acad. Sci. USA (2009).
\bibitem{mermin79} N.D. Mermin, Rev. Mod. Phys. {\bf 51}, 591 (1979).
\bibitem{kleman_book} M, Kl\'{e}man, {\sl Points, lines and walls: in liquid crystals, magnetic systems, and various ordered media} (John Wiley \& Sons; New Jersey; 1983)
\bibitem{trebin82} H.R. Trebin, Adv. Phys {\bf 31}, 195 (1982).
\bibitem{chenthesis} B.G.\ Chen, PhD Thesis, {\sl University of Pennsylvania} (2012).
\bibitem{poenaru81} V. Po\'{e}naru, Commun. Math. Phys. {\bf 80}, 127 (1981).
\bibitem{king78} H.C. King, Ann. Math {\bf 107}, 385 (1978).
\bibitem{king80} H.C. King, Invent. Math  {\bf 62}, 1 (1980).
\bibitem{poenaru77} V. Po\'{e}naru and G. Toulouse, J. Phys. (Paris) {\bf 38}, 887 (1977).
\bibitem{aharoni17} H. Aharoni, T. Machon and R.D. Kamien, Phys. Rev. Lett. {\bf 118}, 257801 (2017).
\bibitem{meyer10} C. Meyer, Y. Nastishin and M. Kl\'{e}man, Phys. Rev. E {\bf 82}, 031704 (2010).
\bibitem{achard05} M.F. Achard, M. Kl\'{e}man, Y. A. Nastishin, and H. T. Nguyen, Eur. Phys. J. E {\bf 16}, 37 (2005).
\bibitem{klemenSM} M. Kl\'{e}man and O.D. Lavrentovitch, {\it Soft Matter Physics:
An Introduction}, (Springer-Verlag; New York; 2003).
\bibitem{williams75} C. E. Williams, Philos. Mag. {\bf 32}, 313 (1975).
\bibitem{alexander12} G.P. Alexander, B.G. Chen, E.A. Matsumoto, R.D. Kamien, Rev. Mod. Phys {\bf 84}, 497 (2012).
\bibitem{pevnyi14} M.Y. Pevnyi, J.V. Selinger, and T.J. Sluckin, Phys. Rev. E {\bf 90}, 032507 (2014).
\bibitem{milnorMorse} J.W. Milnor, {\it Morse Theory}, Annals of Mathematics Studies No. 51. Notes by M. Spivak and R. Wells. (Princeton University Press; Princeton, NJ, 1995).
\bibitem{chen13} B.G. Chen, P.J. Ackerman, G.P. Alexander, R.D. Kamien, and I.I. Smalyukh, Phys. Rev. Lett. {\bf 110}, 237801 (2013).
\bibitem{dennis17} M.R. Dennis and B. Bode, J. Phys. A {\bf 50}, 265204 (2017).
\bibitem{kamien16} R.D. Kamien and R.A. Mosna, New J. Phys. {\bf 18}, 053012 (2016).
\bibitem{blanc04} C. Blanc, N. Zuodar, I. Lelidis, M. K\'{e}man, J.L. Martin, Phys. Rev E {\bf 69}, 011705 (2004).
\bibitem{prep} T Machon, H. Aharoni and R.D. Kamien, {\it In preparation}.
\bibitem{kleman78} M. Kl\'{e}man, L. Michel, Journal de Phys. Lett. (Paris), {\bf 39}, 29, (1978)


\end{thebibliography}
\end{document}